\documentclass[conference]{IEEEtran}
\IEEEoverridecommandlockouts

\usepackage[bottom=1.02in,top=0.71in,left=0.64in,right=0.64in]{geometry}
\usepackage{amsmath}
\usepackage{amsfonts}
\usepackage{amssymb}
\usepackage{bm}
\usepackage[ruled,vlined]{algorithm2e}
\usepackage{mathtools}
\usepackage{amsthm}
\usepackage{booktabs}
\usepackage{subfigure}
\usepackage{tabulary}
\usepackage{footnote}
\usepackage[export]{adjustbox}
\usepackage[justification=centering]{caption}
\usepackage{comment}
\usepackage{cases}
\usepackage{graphicx}
\usepackage{cite}
\usepackage{multicol}
\usepackage{xcolor}
\usepackage{stfloats}
\usepackage{cuted}
\usepackage{lipsum}
\usepackage{dsfont}
\usepackage{optidef}
\usepackage{needspace}

\graphicspath{{./}{figures/}{pictures/}}

\title{Data-Driven Batteryless Channel Sounding for Wi-Fi 8-Inspired Downlink MU-MIMO}

 \author{ \text{Muhan Zhang}\IEEEauthorrefmark{1},  \text{Chuqi Zhang}\IEEEauthorrefmark{1},  \text{Qitong Xu}, \text{Zhaoyu Liu}, \text{Liu Cao}, \text{Lyutianyang Zhang}, \text{Ming Gan}

\thanks{This work was supported by the Youth Innovation Talent Project of Guangdong Provincial Universities (Grant No. 2025KQNCX17).}
 
 \thanks{\IEEEauthorrefmark{1}Both authors contributed equally to this work. Muhan Zhang, Chuqi Zhang, Qitong Xu, Zhaoyu Liu and Liu Cao are with the Electrical and Electronic Engineering Program of both City University of Hong Kong (Dongguan), Dongguan, China and City University of Hong Kong, Hong Kong (e-mail:\{72517272,72517099,72517118,72515198, liu.cao\}@cityu-dg.edu.cn).  Lyutianyang Zhang is with the School of Microelectronics and Communication Engineering, Chongqing University, Chongqing, China (email: zhanglyutianyang@cqu.edu.cn). Ming Gan is with Huawei Technologies Co., Ltd, China (email: ming.gan@huawei.com).}}

\begin{document}

\maketitle
\thispagestyle{empty}

\begin{abstract}
Batteryless overlays couple passive throughput to Wi-Fi sounding overhead and channel state information (CSI) aging. This paper investigates channel sounding for ultra-high reliability (UHR) operation in a Wi-Fi 8/IEEE 802.11bn-inspired downlink multi-user multiple-input multiple-output (MU-MIMO) system with a batteryless passive overlay. We optimize the post-sounding transmission interval to maximize the aggregate throughput of the active Wi-Fi and passive links, while jointly accounting for sounding overhead, CSI aging, modulation and coding scheme (MCS), passive attenuation, and passive data rate. A packet-level cross-layer model evaluates the cycle-average throughput, and a data-driven search identifies the optimal interval under different operating conditions. Simulations demonstrate that passive overlay reshapes the conventional sounding tradeoff: depending on the MCS and passive-link configuration, the additional passive throughput may or may not compensate for the associated Wi-Fi reliability loss, causing the optimal interval to shift. The results provide design guidance for reliable and low-power MU-MIMO WLANs.
\end{abstract}


\section{Introduction}
\label{sec:introduction}

\IEEEPARstart{N}{ext-generation} Wi-Fi is evolving from the
high-efficiency mechanisms of IEEE 802.11ax toward the Extremely High
Throughput (EHT) and Ultra-High Reliability (UHR) objectives of
Wi-Fi 7 and Wi-Fi 8
\cite{ieee80211ax2021,khorov2019tutorial,cao2025revisiting,
galati2024wifi8}. Downlink multi-user multiple-input multiple-output (DL MU-MIMO) is a key technology in this evolution,
as it allows an access point (AP) to simultaneously serve multiple
stations (STAs) through spatial multiplexing. Its performance, however, depends strongly on accurate channel state information
(CSI). The AP must periodically perform channel sounding, collect
compressed beamforming feedback, and construct precoders for the
subsequent downlink transmissions.

The channel-sounding interval introduces an inherent cross-layer
tradeoff. Frequent sounding provides fresh CSI and suppresses
inter-user interference, but consumes a non-negligible fraction of
the available airtime
\cite{zhang2023sounding,sangdeh2021deepmux}. Extending the
post-sounding transmission interval reduces this MAC-layer overhead,
while making the precoder increasingly inconsistent with the
time-varying channel. Such CSI aging degrades the desired signal and
increases residual inter-stream and inter-user interference
\cite{truong2013aging}. Recent work has consequently formulated
Wi-Fi sounding-period selection as a cross-layer optimization problem
that balances MAC-layer CSI overhead against PHY-layer throughput
degradation \cite{zhang2025crosslayer}. Nevertheless, these studies
consider conventional active Wi-Fi transmission without an additional
batteryless service overlaid on the host waveform.

Batteryless communication provides a promising mechanism for
connecting low-cost and energy-constrained devices by reusing existing
RF transmissions. Ambient backscatter first demonstrated
communication without a dedicated active carrier
\cite{liu2013ambient}, while Wi-Fi backscatter and Passive Wi-Fi
further enabled low-power connectivity using Wi-Fi infrastructure
\cite{kellogg2014wifibackscatter,cao2025lightweightcoordinateconditioneddiffusionapproach,gu2025matthew}. Different
from conventional backscatter, Glaze embeds a low-rate downlink
message into an occupied host signal through controlled amplitude
attenuation \cite{kapetanovic2019glaze}. The receiver then decodes the original packet using coherent processing, whereas a
batteryless receiver distinguishes the embedded amplitude states using a low-complexity envelope detector.

Introducing such a passive overlay into a periodically sounded
DL MU-MIMO system fundamentally changes the sounding tradeoff \cite{zhang2025crosslayer}. A
larger attenuation depth improves passive detection but may degrade
the host Wi-Fi packet reliability. The passive embedding rate affects
both the envelope-detection performance and whether the passive
packet fits within the Wi-Fi data field. Moreover, these effects
interact with the Wi-Fi modulation and coding scheme (MCS) and the
CSI aging accumulated during the post-sounding interval. The passive
overlay can therefore shift the throughput-optimal sounding interval
rather than simply contributing a constant additional rate. This
coupled interaction between Glaze-assisted passive transmission and
Wi-Fi sounding-interval optimization remains insufficiently studied.

This paper proposes a batteryless-enabled, Wi-Fi 8-inspired DL MU-MIMO system in which the AP performs EHT-compatible sounding and
then transmits beamformed PPDUs. We build
a packet-level cross-layer model for sounding overhead, outdated precoding, Wi-Fi and passive packet errors, and passive embedding
feasibility, and optimize the post-sounding interval. The main contributions are:

\begin{itemize}
    \item We establish a unified MU-MIMO sounding architecture with
    Glaze amplitude overlay, coherent Wi-Fi reception, and passive
    envelope detection.
    \item We formulate a packet-level cycle-average throughput that
    combines sounding overhead, Wi-Fi goodput, and batteryless
    goodput, and optimize \(T_{\mathrm p}\) under each fixed
    operating condition.
    \item We characterize overhead-limited and aging-limited regimes
    and give a linear-complexity data-driven search for the
    throughput-optimal sounding interval.
\end{itemize}

The remainder of this paper presents the system model and
sounding-interval formulation in Section~II, link-level simulations
in Section~III, and conclusions in Section~IV.

\section{System Model and Sounding Optimization}
\label{sec:sys_model}

\begin{figure*}[!t]
    \centering
    \includegraphics[width=0.85\textwidth, trim=0 445 0 10, clip]{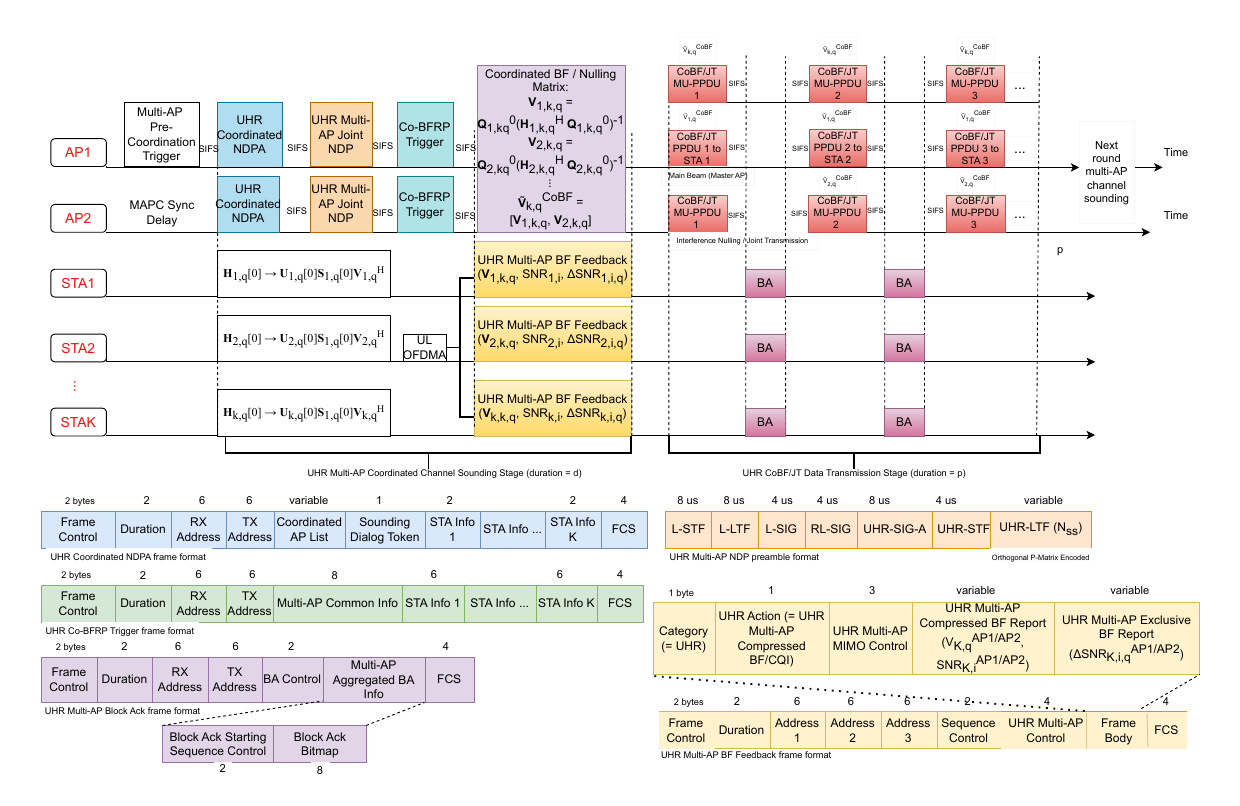}
    \caption{Procedure of UHR Channel Sounding and MU-MIMO Transmission in Wi-Fi 8 Systems.}
    \label{fig:system_architecture}
    \vspace{-0.5cm}
\end{figure*}

\subsection{Wi-Fi 8 Channel Sounding and Transmission}

As shown in Fig.~\ref{fig:system_architecture}, the system is governed
by a cross-layer tradeoff between CSI freshness and sounding overhead.
A short post-sounding interval improves beamforming accuracy but
spends more airtime on sounding, whereas a long interval reduces
overhead but increases residual MU-MIMO interference. The passive
overlay further changes this balance because its attenuation depth and
embedding rate affect both Wi-Fi reliability and batteryless throughput. An AP with \(N_{\mathrm t}\) antennas serves active STAs
\(\mathcal{K}=\{1,\ldots,K\}\). STA \(k\) has
\(N_{\mathrm r,k}\) receive antennas and \(N_{\mathrm s,k}\)
spatial streams, and a batteryless receiver observes the same
downlink waveform. Each cycle contains sounding with duration
\(T_{\mathrm d}\) and a post-sounding data interval
\(T_{\mathrm p}\):
\begin{equation}
T_{\mathrm{cyc}}=T_{\mathrm d}+T_{\mathrm p}.
\label{eq:cycle_duration}
\end{equation}

The AP transmits EHT-compatible NDPA and NDP frames, polls
beamforming feedback, and receives compressed CSI from the scheduled
STAs. With one BFRP round, the sounding duration is
\begin{equation}
\begin{aligned}
T_{\mathrm d}={}&T_{\mathrm{access}}+T_{\mathrm{NDPA}}
+T_{\mathrm{NDP}}+T_{\mathrm{BFRP}}+T_{\mathrm{BFFB}}\\
&+N_{\mathrm{SIFS}}T_{\mathrm{SIFS}} .
\end{aligned}
\label{eq:sounding_duration}
\end{equation}

Let \(\mathbf{H}_{k,q}[0]\) be the channel of STA \(k\) on
subcarrier \(q\) at the sounding instant. The STA feeds back a
compressed representation of the dominant right singular vectors:
\begin{equation}
\mathbf{H}_{k,q}[0]
=\mathbf{U}_{k,q}[0]\boldsymbol{\Sigma}_{k,q}[0]
\mathbf{V}_{k,q}^{H}[0].
\label{eq:sounding_svd}
\end{equation}
For block-diagonalization precoding, the AP concatenates the feedback
of the other users into \(\overline{\mathbf{V}}_{k,q}\), finds a
null-space basis \(\mathbf{Q}_{k,q}^{(0)}\), and sets
\begin{equation}
\widetilde{\mathbf{V}}_{k,q}
=\mathbf{Q}_{k,q}^{(0)}
\left(\mathbf{V}_{k,q}^{H}\mathbf{Q}_{k,q}^{(0)}\right)^{\dagger}
\mathbf{D}_{k,q},
\label{eq:bd_precoder}
\end{equation}
where \(\mathbf{D}_{k,q}\) normalizes power. The precoder is fixed
during \(T_{\mathrm p}\), so channel evolution gradually weakens the
interference-nulling condition.

\subsection{Time-Varying Channel Model}
\label{subsec:channel_model}

The initial frequency-domain channel follows an indoor multipath
model:
\begin{equation}
\mathbf{H}_{k,q}[0]
=\sum_{\ell=1}^{L_k}
\mathbf{A}_{k,\ell}\exp(-j2\pi f_q\tau_{k,\ell}),
\label{eq:initial_channel}
\end{equation}
where \(\mathbf{A}_{k,\ell}\) and \(\tau_{k,\ell}\) are the MIMO
tap matrix and delay. CSI aging is modeled by
\begin{equation}
\mathbf{H}_{k,q}[n]
=\rho_{k,q}\mathbf{H}_{k,q}[n-1]
+\sqrt{1-|\rho_{k,q}|^{2}}\mathbf{E}_{k,q}[n],
\label{eq:ar_channel}
\end{equation}
where \(\mathbf{E}_{k,q}[n]\) is an independent complex Gaussian
innovation. Under isotropic scattering,
\(\rho_{k,q}=J_0(2\pi f_{\mathrm D,k}T_{\mathrm s})\) and
\(f_{\mathrm D,k}=v_kf_{\mathrm c}/c\). Conditioned on the sounded
channel, the prediction mismatch grows as
\begin{equation}
\begin{aligned}
&\mathbb{E}\!\left[
\left\|\mathbf{H}_{k,q}[n]-\rho_{k,q}^{n}
\mathbf{H}_{k,q}[0]\right\|_{\mathrm F}^{2}\right]\\
&\quad
=N_{\mathrm r,k}N_{\mathrm t}\sigma_{H,k,q}^{2}
\left(1-|\rho_{k,q}|^{2n}\right).
\end{aligned}
\label{eq:channel_mismatch}
\end{equation}

\subsection{PHY-Layer Glaze-Assisted Transmission}
\label{subsec:phy_glaze}

Before passive overlay, the MU-MIMO signal on subcarrier \(q\) is
\begin{equation}
\mathbf{x}_{q}[n]
=\sum_{k=1}^{K}\sqrt{\frac{P_k[n]}{N_{\mathrm s,k}}}
\widetilde{\mathbf{V}}_{k,q}\mathbf{s}_{k,q}[n],
\qquad
\sum_{k=1}^{K}P_k[n]\le P_{\max}.
\label{eq:host_mu_signal}
\end{equation}
The passive bits are Manchester encoded. For attenuation depth
\(\Delta\), the two amplitude states are
\begin{equation}
a(b_{\ell};\Delta)=
\begin{cases}
10^{-\Delta/20},& b_{\ell}=0,\\
1,& b_{\ell}=1 .
\end{cases}
\label{eq:glaze_attenuation}
\end{equation}
A packet gate \(g_n(t)\) enables overlay only in eligible Wi-Fi data
fields:
\begin{equation}
\widetilde{\mathbf{x}}_{q}[n,t]
=\left[1-g_n(t)+g_n(t)a(b_{\ell};\Delta)\right]
\mathbf{x}_{q}[n,t].
\label{eq:gated_overlay_signal}
\end{equation}
Because the same scalar is applied across antennas and streams, the
overlay changes the waveform amplitude but not the spatial
beamforming direction. For equiprobable states,
\(\overline{a}=(1+a_{\Delta})/2\) and
\(\sigma_a^2=(1-a_{\Delta})^2/4\).

The active Wi-Fi STA still performs coherent MIMO reception. Its
received signal on subcarrier \(q\) is
\begin{equation}
\begin{aligned}
\mathbf{y}_{k,q}[n]
={}&a(b_{\ell};\Delta)\mathbf{H}_{k,q}[n]
\sum_{j=1}^{K}\sqrt{\frac{P_j[n]}{N_{\mathrm s,j}}}
\widetilde{\mathbf{V}}_{j,q}\mathbf{s}_{j,q}[n]\\
&+\mathbf{z}_{k,q}[n],
\end{aligned}
\label{eq:wifi_received_signal}
\end{equation}
where \(\mathbf{z}_{k,q}[n]\) is receiver noise. With combining
vector \(\mathbf{u}_{k,i,q}[n]\), the desired power of stream \(i\)
is
\begin{equation}
S_{k,i,q}[n]=
\frac{P_k[n]}{N_{\mathrm s,k}}
\left|
\mathbf{u}_{k,i,q}^{H}[n]\mathbf{H}_{k,q}[n]
\widetilde{\mathbf{v}}_{k,i,q}
\right|^{2},
\end{equation}
and the residual intra-user plus inter-user interference is
\begin{equation}
\begin{aligned}
I_{k,i,q}[n]
={}&
\sum_{\substack{r=1\\r\neq i}}^{N_{\mathrm s,k}}
\frac{P_k[n]}{N_{\mathrm s,k}}
\left|
\mathbf{u}_{k,i,q}^{H}[n]\mathbf{H}_{k,q}[n]
\widetilde{\mathbf{v}}_{k,r,q}
\right|^{2}\\
&+
\sum_{\substack{j=1\\j\neq k}}^{K}
\sum_{r=1}^{N_{\mathrm s,j}}
\frac{P_j[n]}{N_{\mathrm s,j}}
\left|
\mathbf{u}_{k,i,q}^{H}[n]\mathbf{H}_{k,q}[n]
\widetilde{\mathbf{v}}_{j,r,q}
\right|^{2}.
\end{aligned}
\end{equation}
These terms are small just after sounding but increase as the fixed
precoder becomes outdated. The zero-mean amplitude switching is
modeled as
\[
D_{k,i,q}^{\mathrm{ovl}}[n]
=\xi_{\mathrm W}(R_{\mathrm b},m_k)\sigma_a^2
\left(S_{k,i,q}[n]+I_{k,i,q}[n]\right),
\]
which gives the effective Wi-Fi SINR
\begin{equation}
\gamma_{k,i,q}^{\mathrm W}[n]
=\frac{\overline{a}^{\,2}S_{k,i,q}[n]}
{\overline{a}^{\,2}I_{k,i,q}[n]
+D_{k,i,q}^{\mathrm{ovl}}[n]
+\sigma_{k,q}^{2}\|\mathbf{u}_{k,i,q}[n]\|_{2}^{2}}.
\label{eq:wifi_distortion_sinr}
\end{equation}
The packet error probability is abstracted as
\begin{equation}
P_{e,k}^{\mathrm W}[n]
=f_{\mathrm W}\!\left(
\gamma_{k,\mathrm{eff}}^{\mathrm W}[n],m_k,L_k^{\mathrm W}
\right).
\label{eq:wifi_per}
\end{equation}

The batteryless receiver performs envelope detection rather than
coherent MIMO processing. After impedance matching, rectification,
integration, and threshold comparison, a scalar-equivalent model is
\begin{equation}
r^{\mathrm P}[n,t]
=g^{\mathrm P}[n]a(b_{\ell};\Delta)x[n,t]
+v^{\mathrm P}[n,t],
\end{equation}
\begin{equation}
z^{\mathrm P}[n,t]=|r^{\mathrm P}[n,t]|+\eta[n,t],
\label{eq:passive_observation}
\end{equation}
where \(g^{\mathrm P}[n]\) is the equivalent AP-to-passive channel,
\(x[n,t]\) is the scalar host waveform, and \(\eta[n,t]\) models
front-end and comparator distortion. With equivalent incident power
\(\Omega^{\mathrm P}[n]=\mathbb{E}[|g^{\mathrm P}[n]x[n,t]|^2]\),
the passive detection SNR is
\begin{equation}
\gamma^{\mathrm P}[n]
=\frac{F_{\mathrm s}(1-a_{\Delta})^{2}\Omega^{\mathrm P}[n]}
{R_{\mathrm b}\sigma_{\mathrm P}^{2}},
\label{eq:passive_detection_snr}
\end{equation}
and
\begin{equation}
P_b^{\mathrm P}[n]=f_{\mathrm P}(\gamma^{\mathrm P}[n]),\qquad
P_e^{\mathrm P}[n]
=1-\left(1-P_b^{\mathrm P}[n]\right)^{K_{\mathrm P}}.
\label{eq:passive_per}
\end{equation}
Increasing \(\Delta\) improves passive separation but can reduce
Wi-Fi reliability; increasing \(R_{\mathrm b}\) shortens each
passive-bit observation window.

For MCS vector \(\mathbf{m}\), the Wi-Fi PHY rate and common data
duration are
\begin{equation}
\begin{aligned}
R_{\mathrm W,k}(m_k)
&=\frac{N_{\mathrm{SD}}N_{\mathrm{BPSC}}(m_k)r_{\mathrm c}(m_k)
N_{\mathrm s,k}}{T_{\mathrm{DFT}}+T_{\mathrm{GI}}},\\
T_{\mathrm{data}}(\mathbf{m})
&=\max_{k\in\mathcal K}\frac{L_k^{\mathrm W}}{R_{\mathrm W,k}(m_k)}.
\end{aligned}
\label{eq:wifi_phy_rate}
\end{equation}
Passive embedding is feasible only if
\begin{equation}
\chi^{\mathrm{fit}}(R_{\mathrm b},\mathbf{m})
=\mathbf{1}\!\left\{
T_{\mathrm{pre}}^{\mathrm P}+\frac{K_{\mathrm c}}{R_{\mathrm b}}
\le T_{\mathrm{data}}(\mathbf{m})\right\}.
\label{eq:passive_fit}
\end{equation}
We collect the fixed operating condition as
\(\boldsymbol{\theta}=(\Delta,R_{\mathrm b},\mathbf{m})\); the
primary optimization variable is \(T_{\mathrm p}\).

\subsection{MAC-Layer Sounding Overhead and Cycle Throughput}
\label{subsec:mac_throughput}

For a fixed \(\boldsymbol{\theta}\), one post-sounding transmission
unit has duration
\begin{equation}
\begin{aligned}
T_0(\boldsymbol{\theta})
&=T_{\mathrm{pre}}^{\mathrm W}+T_{\mathrm{data}}(\mathbf{m})
+T_{\mathrm{gap}},\\
T_{\mathrm p}
&=pT_0(\boldsymbol{\theta}),\\
d(\boldsymbol{\theta})
&=\frac{T_{\mathrm d}}{T_0(\boldsymbol{\theta})}.
\end{aligned}
\label{eq:constant_goodput}
\end{equation}
Here \(T_{\mathrm{gap}}\) includes per-packet frame-exchange
overhead, and \(d(\boldsymbol{\theta})\) is the normalized sounding
overhead.

The expected delivered Wi-Fi and passive payload bits in transmission
unit \(n\) are
\begin{equation}
\begin{aligned}
D_{\mathrm W}[n;\boldsymbol{\theta}]
&=\sum_{k=1}^{K}L_k^{\mathrm W}
\left(1-P_{e,k}^{\mathrm W}[n;\boldsymbol{\theta}]\right),\\
D_{\mathrm P}[n;\boldsymbol{\theta}]
&=K_{\mathrm P}\chi^{\mathrm{fit}}(\boldsymbol{\theta})
\left(1-P_e^{\mathrm P}[n;\boldsymbol{\theta}]\right).
\end{aligned}
\end{equation}
The aggregate packet-level throughput is
\begin{equation}
z_{\mathrm{tot}}[n;\boldsymbol{\theta}]
=\frac{D_{\mathrm W}[n;\boldsymbol{\theta}]
+\lambda_{\mathrm P}D_{\mathrm P}[n;\boldsymbol{\theta}]}
{T_0(\boldsymbol{\theta})},
\label{eq:instantaneous_goodput}
\end{equation}
where \(\lambda_{\mathrm P}=1\) unless otherwise specified. The
cycle-average aggregate throughput is
\begin{equation}
\begin{aligned}
\overline{R}_{\mathrm{sum}}(p;\boldsymbol{\theta})
&=\frac{\sum_{n=1}^{p}\mathbb{E}\!\left[
D_{\mathrm W}[n;\boldsymbol{\theta}]
+\lambda_{\mathrm P}D_{\mathrm P}[n;\boldsymbol{\theta}]
\right]}{T_{\mathrm d}+pT_0(\boldsymbol{\theta})}
\\
&=\frac{\sum_{n=1}^{p}\mathbb{E}\!\left[
z_{\mathrm{tot}}[n;\boldsymbol{\theta}]\right]}
{d(\boldsymbol{\theta})+p}.
\end{aligned}
\label{eq:unified_cycle_throughput}
\end{equation}

\subsection{Cross-Layer Sounding-Interval Optimization}
\label{subsec:sounding_optimization}

\textbf{Problem 1 (Batteryless Sounding-Interval Optimization):}
For fixed \(\boldsymbol{\theta}\), find
\begin{equation}
p^{\star}(\boldsymbol{\theta})
=\underset{1\le p\le P_{\max}}{\operatorname{arg\,max}}\;
\overline{R}_{\mathrm{sum}}(p;\boldsymbol{\theta}),\qquad
T_{\mathrm p}^{\star}(\boldsymbol{\theta})
=p^{\star}(\boldsymbol{\theta})T_0(\boldsymbol{\theta}).
\end{equation}
The feasible operating set enforces Wi-Fi reliability, passive
reliability, and embedding duration:
\begin{equation}
\Theta_{\mathrm{fea}}
=\left\{\boldsymbol{\theta}:
\begin{array}{l}
P_{e,k}^{\mathrm W}[n;\boldsymbol{\theta}]\le\epsilon_{\mathrm W},
\ \forall k,n,\\
P_e^{\mathrm P}[n;\boldsymbol{\theta}]\le\epsilon_{\mathrm P},
\ \forall n,\\
T_{\mathrm{pre}}^{\mathrm P}+K_{\mathrm c}/R_{\mathrm b}
\le T_{\mathrm{data}}(\mathbf{m})
\end{array}\right\}.
\label{eq:feasible_operating_set}
\end{equation}
The search is repeated over MCS levels and passive-overlay
configurations to obtain the performance envelope.

\subsection{Theoretical Analysis of the Sounding Tradeoff}
\label{subsec:theoretical_analysis}

In an overhead-limited regime, the expected packet throughput is nearly
constant:
\(\mathbb{E}[z_{\mathrm{tot}}[n;\boldsymbol{\theta}]]
=C_{\boldsymbol{\theta}}\) for
\(n=1,\ldots,P_{\max}\). Then
\begin{equation}
\begin{aligned}
\overline{R}_{\mathrm{sum}}(p;\boldsymbol{\theta})
&=\frac{p}{d(\boldsymbol{\theta})+p}C_{\boldsymbol{\theta}},\\
\overline{R}_{\mathrm{sum}}(p+1;\boldsymbol{\theta})
-\overline{R}_{\mathrm{sum}}(p;\boldsymbol{\theta})
&{}\\[-1.5ex]
&=\frac{C_{\boldsymbol{\theta}}d(\boldsymbol{\theta})}
{[d(\boldsymbol{\theta})+p+1][d(\boldsymbol{\theta})+p]}
\ge0.
\end{aligned}
\end{equation}
Thus no interior optimum exists and
\begin{equation}
p^{\star}(\boldsymbol{\theta})=P_{\max}.
\label{eq:block_fading_optimum}
\end{equation}
This explains the low-MCS regime where robust modulation keeps the
throughput profile nearly flat.

With channel aging, define
\[
A_{\boldsymbol{\theta}}(p)
=\sum_{n=1}^{p}\mathbb{E}\!\left[
z_{\mathrm{tot}}[n;\boldsymbol{\theta}]\right].
\]
The one-step difference becomes
\begin{equation}
\overline{R}_{\mathrm{sum}}(p+1;\boldsymbol{\theta})
-\overline{R}_{\mathrm{sum}}(p;\boldsymbol{\theta})
=\frac{\mathbb{E}\!\left[
z_{\mathrm{tot}}[p+1;\boldsymbol{\theta}]\right]
-\overline{R}_{\mathrm{sum}}(p;\boldsymbol{\theta})}
{d(\boldsymbol{\theta})+p+1}.
\label{eq:one_step_difference}
\end{equation}
Therefore,
\begin{equation}
\begin{aligned}
\overline{R}_{\mathrm{sum}}(p+1;\boldsymbol{\theta})
&\ge\overline{R}_{\mathrm{sum}}(p;\boldsymbol{\theta})\\
&\Longleftrightarrow
\mathbb{E}\!\left[z_{\mathrm{tot}}[p+1;\boldsymbol{\theta}]\right]
\ge\overline{R}_{\mathrm{sum}}(p;\boldsymbol{\theta}).
\end{aligned}
\label{eq:increase_condition}
\end{equation}
Extending the interval is beneficial only while the next packet's
expected throughput exceeds the current cycle average.

The same first-difference expression also explains why the objective
is typically unimodal. Suppose the expected packet-level throughput is
non-increasing with channel age, i.e.,
\begin{equation}
\mathbb{E}[z_{\mathrm{tot}}[1;\boldsymbol{\theta}]]
\ge \cdots \ge
\mathbb{E}[z_{\mathrm{tot}}[P_{\max};\boldsymbol{\theta}]]
\ge0.
\end{equation}
If the cycle average first stops increasing at index \(p\), then
\(\mathbb{E}[z_{\mathrm{tot}}[p+1;\boldsymbol{\theta}]]
\le \overline{R}_{\mathrm{sum}}(p;\boldsymbol{\theta})\). Hence
\begin{equation}
\begin{aligned}
\overline{R}_{\mathrm{sum}}(p+1;\boldsymbol{\theta})
&=\frac{(d+p)\overline{R}_{\mathrm{sum}}(p;\boldsymbol{\theta})
+\mathbb{E}[z_{\mathrm{tot}}[p+1;\boldsymbol{\theta}]]}{d+p+1}\\
&\ge \mathbb{E}[z_{\mathrm{tot}}[p+1;\boldsymbol{\theta}]]
\ge \mathbb{E}[z_{\mathrm{tot}}[p+2;\boldsymbol{\theta}]] .
\end{aligned}
\end{equation}
Substituting this inequality back into
\eqref{eq:one_step_difference} shows that
\(\overline{R}_{\mathrm{sum}}(p+2;\boldsymbol{\theta})
\le\overline{R}_{\mathrm{sum}}(p+1;\boldsymbol{\theta})\). Repeating
the argument gives a non-increasing tail after the first decreasing
point. Thus, in the ideal monotone-aging case, the optimal sounding
interval is the first local maximum; in packet-level simulations,
finite-sample fluctuations can slightly perturb this shape, so the
implementation still evaluates the full candidate range.

It is also useful to separate the passive-enabled throughput into the
legacy Wi-Fi part and the passive contribution. Let
\[
z_{\mathrm{tot}}[n;\boldsymbol{\theta}]
=z_{\mathrm W}[n;\boldsymbol{\theta}]
+\lambda_{\mathrm P}z_{\mathrm P}[n;\boldsymbol{\theta}],
\]
where \(z_{\mathrm W}\) already includes the reliability loss induced
by attenuation and amplitude switching. Relative to the no-overlay
baseline \(z_{\mathrm B}[n]\), the instantaneous gain can be written
as
\begin{equation}
\Delta z[n;\boldsymbol{\theta}]
=z_{\mathrm{tot}}[n;\boldsymbol{\theta}]-z_{\mathrm B}[n]
=\lambda_{\mathrm P}z_{\mathrm P}[n;\boldsymbol{\theta}]
-\delta_{\mathrm W}[n;\boldsymbol{\theta}],
\label{eq:passive_gain_decomposition}
\end{equation}
where \(\delta_{\mathrm W}\ge0\) denotes the Wi-Fi throughput loss
caused by passive overlay. A larger attenuation depth \(\Delta\)
typically increases \(z_{\mathrm P}\) by improving envelope
separation, but it also increases \(\delta_{\mathrm W}\) through
lower effective SINR. Similarly, a larger passive rate
\(R_{\mathrm b}\) may raise the nominal passive bit rate, but it
reduces the number of envelope samples per bit and can increase
passive packet errors.

The direction in which passive overlay shifts the optimal sounding
interval follows directly from the one-step condition. Let
\(\overline{R}_{\mathrm B}(p)\) denote the baseline cycle average.
At the baseline optimum \(p_{\mathrm B}^{\star}\), the passive-enabled
system prefers a shorter interval if
\begin{equation}
\mathbb{E}\!\left[
z_{\mathrm{tot}}[p_{\mathrm B}^{\star};\boldsymbol{\theta}]
\right]
<\overline{R}_{\mathrm{sum}}
(p_{\mathrm B}^{\star}-1;\boldsymbol{\theta}),
\label{eq:left_shift_condition}
\end{equation}
because the last packets in the cycle no longer justify the reduced
sounding overhead. This case occurs when the Wi-Fi loss
\(\delta_{\mathrm W}\) is strongest at aged CSI states, for example
under high MCS or deep attenuation. Conversely, if the passive
throughput remains robust over the later packets and
\(\Delta z[n;\boldsymbol{\theta}]\) is positive over a wide range of
\(n\), then the passive service increases the marginal value of
longer cycles and may shift \(p^{\star}\) to the right. When the two
effects are nearly balanced, the optimum remains close to the
baseline value. These three cases correspond to the left-shift,
right-shift, and nearly unchanged behaviors observed in the
simulation results.

\subsection{Data-Driven Sounding-Interval Search}
\label{subsec:data_driven_search}

Because the expectation in
\eqref{eq:unified_cycle_throughput} is unavailable in closed form, we
estimate it from \(N_{\mathrm R}\) independent channel realizations:
\begin{equation}
\widehat{z}_{\mathrm{tot}}[n;\boldsymbol{\theta}]
=\frac{1}{N_{\mathrm R}}\sum_{r=1}^{N_{\mathrm R}}
z_{\mathrm{tot}}^{(r)}[n;\boldsymbol{\theta}].
\label{eq:empirical_goodput}
\end{equation}
For a candidate \(p\),
\begin{equation}
\widehat{A}_{\boldsymbol{\theta}}(p)
=\sum_{n=1}^{p}\widehat{z}_{\mathrm{tot}}[n;\boldsymbol{\theta}],
\qquad
\widehat{R}_{\mathrm{sum}}(p;\boldsymbol{\theta})
=\frac{\widehat{A}_{\boldsymbol{\theta}}(p)}
{d(\boldsymbol{\theta})+p}.
\label{eq:empirical_cycle_throughput}
\end{equation}
The empirical solution is
\begin{equation}
\widehat{p}^{\star}(\boldsymbol{\theta})
=\underset{1\le p\le P_{\max}}{\operatorname{arg\,max}}\;
\widehat{R}_{\mathrm{sum}}(p;\boldsymbol{\theta}),\qquad
\widehat{T}_{\mathrm p}^{\star}(\boldsymbol{\theta})
=\widehat{p}^{\star}(\boldsymbol{\theta})T_0(\boldsymbol{\theta}).
\label{eq:data_driven_solution}
\end{equation}
The accumulated throughput is updated recursively as
\(\widehat{A}_{\boldsymbol{\theta}}(p)
=\widehat{A}_{\boldsymbol{\theta}}(p-1)
+\widehat{z}_{\mathrm{tot}}[p;\boldsymbol{\theta}]\), so evaluating
all candidates requires \(\mathcal{O}(P_{\max})\) operations. The
complete search is summarized in Algorithm~\ref{alg:data_driven_search}.
In practice, each realization records the decoded Wi-Fi payload,
passive decoding result, packet duration, and sounding overhead under
the same operating condition \(\boldsymbol{\theta}\). Averaging over
realizations therefore preserves the packet ordering within a
sounding round while smoothing small-scale fading randomness. The
training and validation sets used in Section~\ref{sec:simulations}
are generated independently, so agreement between their optima
indicates that the selected \(T_{\mathrm p}^{\star}\) reflects the
ensemble tradeoff rather than a realization-specific fading event.

\begin{algorithm}[t]
\caption{Data-Driven Optimal Sounding-Interval Search}
\label{alg:data_driven_search}

\SetKwInOut{Input}{Input}
\SetKwInOut{Output}{Output}
\SetKwComment{Comment}{// }{}

\Input{
Empirical aggregate-goodput profile
\(\left\{
\widehat{z}_{\mathrm{tot}}[n;\bm{\theta}]
\right\}_{n=1}^{P_{\max}}\),
normalized sounding overhead \(d(\bm{\theta})\), and
packet-level duration \(T_0(\bm{\theta})\)
}

\Output{
Optimal packet count
\(\widehat{p}^{\star}(\bm{\theta})\)
and optimal post-sounding interval
\(\widehat{T}_{\mathrm p}^{\star}(\bm{\theta})\)
}

\BlankLine

Initialize
\(\widehat{A}_{\bm{\theta}}\leftarrow 0\),
\(\widehat{R}_{\max}\leftarrow -\infty\), and
\(\widehat{p}^{\star}(\bm{\theta})\leftarrow 1\)\;

\For{\(p\leftarrow 1\) \KwTo \(P_{\max}\)}{

    \(\widehat{A}_{\bm{\theta}}
    \leftarrow
    \widehat{A}_{\bm{\theta}}
    +
    \widehat{z}_{\mathrm{tot}}[p;\bm{\theta}]\)\;

    \(\widehat{R}_{\mathrm{sum}}(p;\bm{\theta})
    \leftarrow
    \dfrac{
    \widehat{A}_{\bm{\theta}}
    }{
    d(\bm{\theta})+p
    }\)\;

    \If{
    \(\widehat{R}_{\mathrm{sum}}(p;\bm{\theta})
    >
    \widehat{R}_{\max}\)
    }{

        \(\widehat{R}_{\max}
        \leftarrow
        \widehat{R}_{\mathrm{sum}}(p;\bm{\theta})\)\;

        \(\widehat{p}^{\star}(\bm{\theta})
        \leftarrow p\)\;
    }
}

\(\widehat{T}_{\mathrm p}^{\star}(\bm{\theta})
\leftarrow
\widehat{p}^{\star}(\bm{\theta})
T_0(\bm{\theta})\)\;

\Return{
\(\widehat{p}^{\star}(\bm{\theta})\),
\(\widehat{T}_{\mathrm p}^{\star}(\bm{\theta})\)
}\;

\end{algorithm}
\section{Simulations}
\label{sec:simulations}

We evaluate the proposed data-driven sounding optimization in a
Wi-Fi~8-inspired UHR WLAN and compare it with fixed-period sounding. The MATLAB link simulator uses an HE/TGax-based MU-MIMO proxy with
TGax Model-D fading. The AP has \(N_t=4\) antennas and serves users
with \(N_r=2\) and \(N_{ss}=2\). The carrier frequency is 5.25 GHz,
the bandwidth is 20 MHz, the mobility speed is 0.089 m/s, and each
sounding round contains 150 packets, corresponding to about 180 ms.
The main settings are listed in Table~\ref{tab:settings}.

\begin{table}[t]
\caption{Simulation Settings}
\label{tab:settings}
\centering
\footnotesize
\setlength{\tabcolsep}{3pt}
\begin{tabular}{|l|l|}
\hline
\textbf{Parameters} & \textbf{Value} \\ \hline
Simulation proxy & HE/TGax link-level model \\ \hline
Scenario & Wi-Fi~8-inspired UHR WLAN \\ \hline
Carrier frequency & 5.25 GHz \\ \hline
Bandwidth & 20 MHz (242 tones) \\ \hline
Channel model & IEEE TGax Model D \\ \hline
Environmental speed & 0.089 m/s \\ \hline
Sounding overhead \(T_d\) & 8 units (\(\approx\) 8 ms) \\ \hline
MCS range & Index \(\{0,\ldots,9\}\) \\ \hline
MPDU length & 1000 bytes \\ \hline
\end{tabular}
\end{table}

\subsection{Performance of the Proposed Sounding Period Optimization}
\begin{figure}[t]
    \centering
    \includegraphics[width=\linewidth]{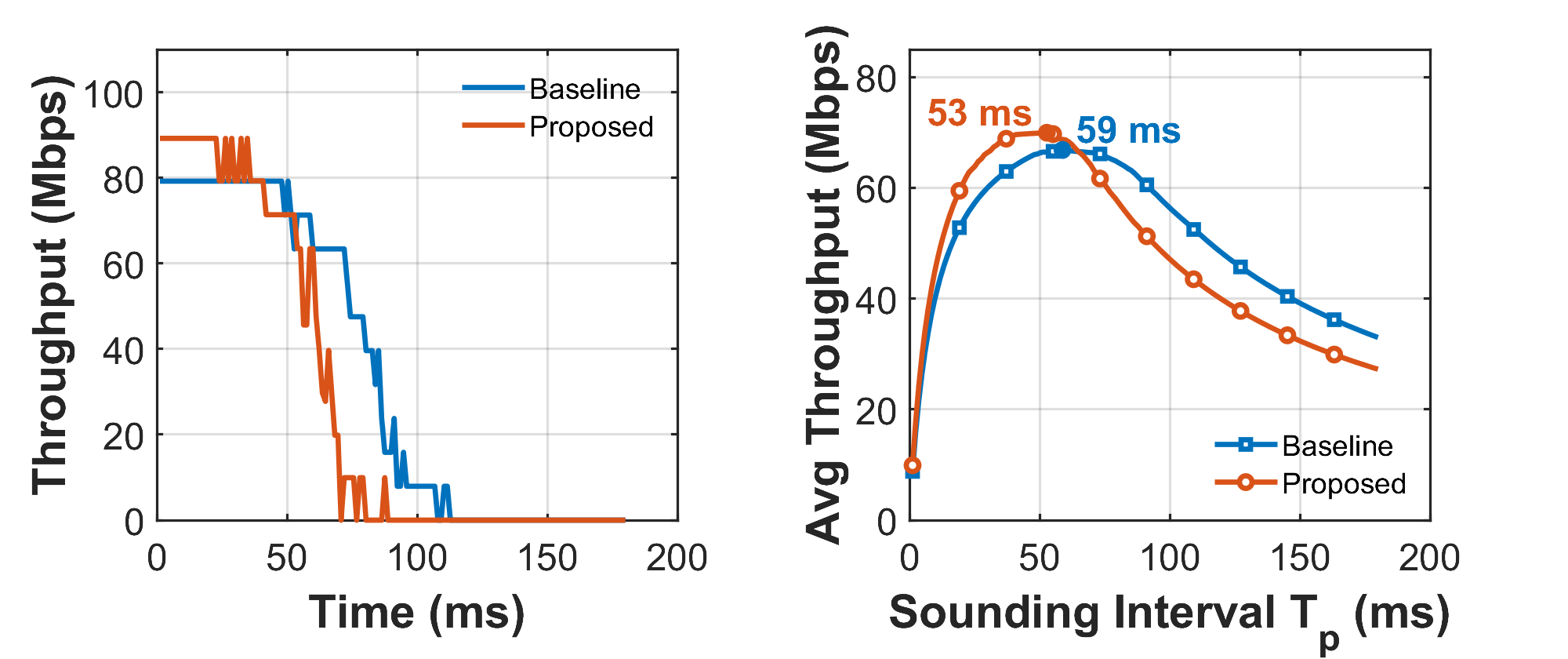}
    \caption{Throughput comparison for MCS 4: instantaneous throughput and cycle-average throughput versus \(T_p\).}
    \label{fig:mcs4}
\end{figure}

Fig.~\ref{fig:mcs4} shows the representative MCS-4 case. With fresh
CSI, the integrated link reaches about 90 Mbps including the
10 Mbps passive gain, but the instantaneous throughput drops to
about 18 Mbps after 150 ms because outdated precoding increases
inter-user interference.

\begin{figure}[t]
    \centering
    \includegraphics[width=0.9\linewidth]{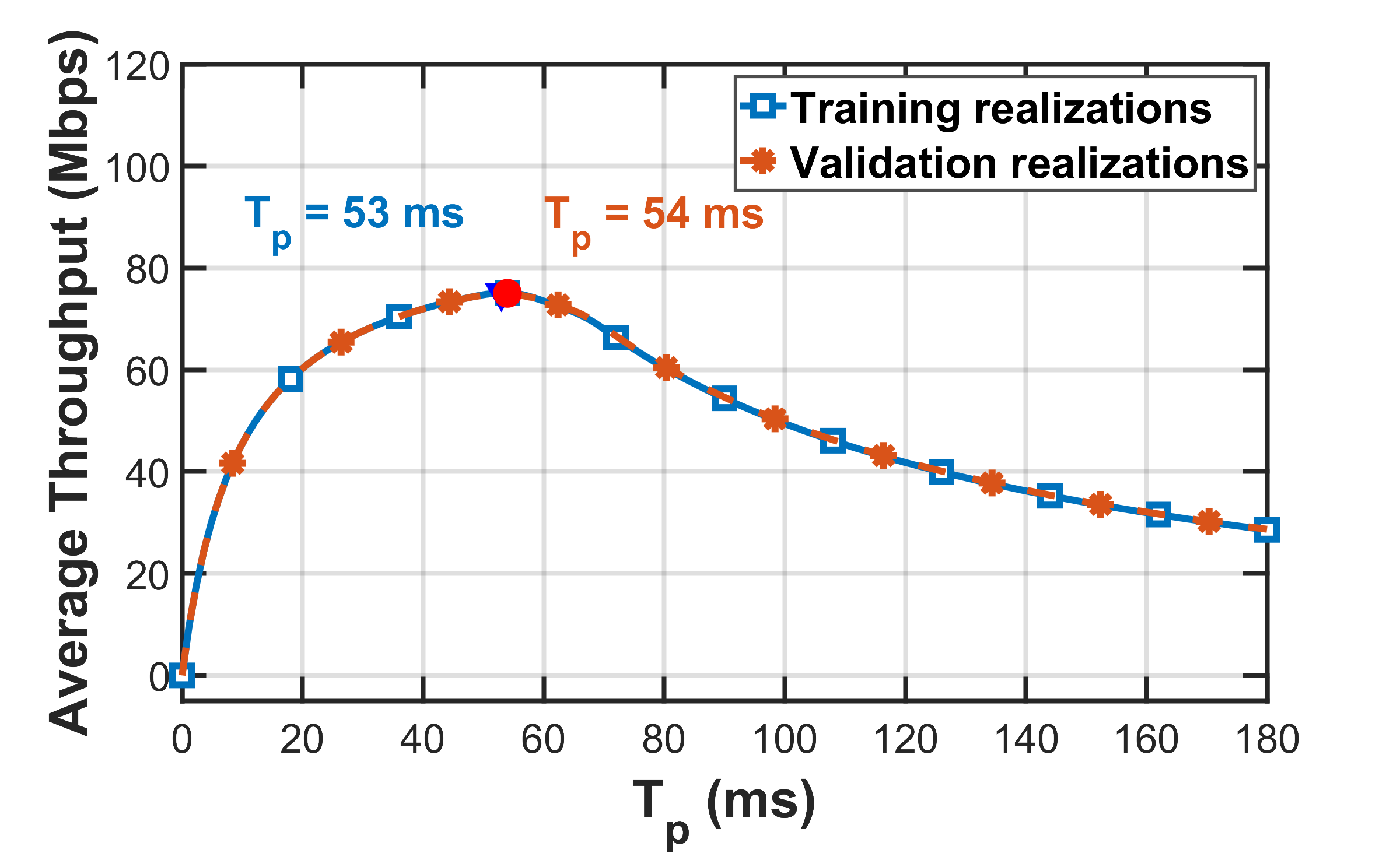}
    \caption{Data-driven validation with training and validation channel realizations.}
    \label{fig:data_driven}
\end{figure}

The average-throughput curve in Fig.~\ref{fig:mcs4} gives
\(T_p^\star=53\) ms for the proposed passive-enabled system and
59 ms for the baseline, indicating that passive-overlay loss can
justify more frequent CSI refresh. Fig.~\ref{fig:data_driven}
validates the data-driven search: the optimum estimated from
\(K=30\) training realizations differs from the \(N=20\) validation
optimum by only 1 ms.

This result also illustrates the non-constant effect of passive
overlay. At the beginning of the cycle, the passive stream contributes
additional payload while the Wi-Fi link still benefits from fresh CSI.
As the channel ages, however, the same attenuation and switching
distortion become more costly because the MU-MIMO interference margin
is already reduced by outdated precoding. The integrated throughput
therefore falls faster than the baseline in the later part of the
sounding round, which explains why the proposed optimum is shifted to
a shorter interval in the MCS-4 case. The small training-validation
gap further suggests that the empirical profile is smooth enough for
practical interval selection even when individual packets experience
different fading realizations.

\subsection{System Performance Envelope across MCS Index}
\begin{figure}[t]
    \centering
    \includegraphics[width=\linewidth]{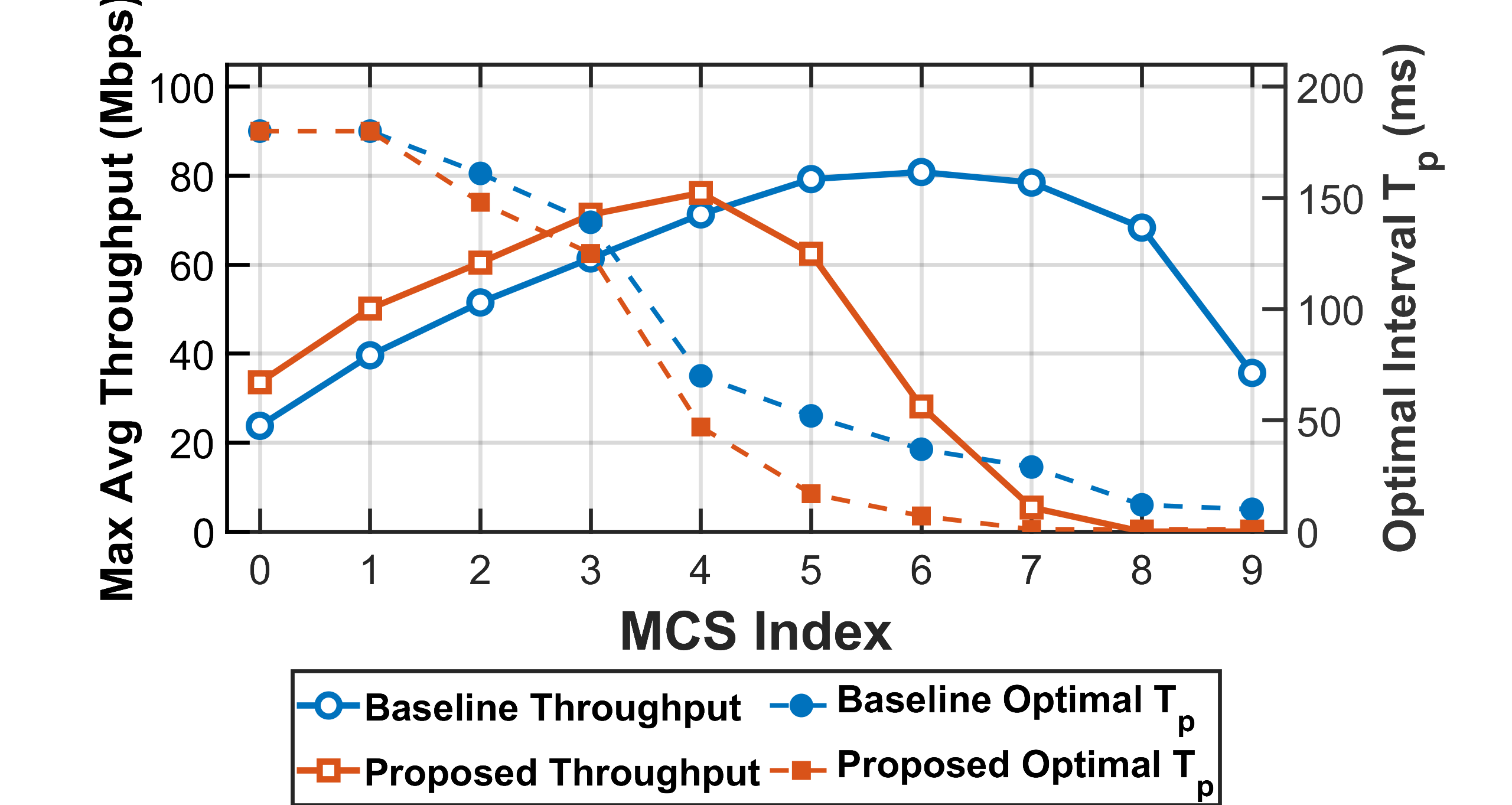}
    \caption{Maximum average throughput and optimal \(T_p^\star\) versus MCS index.}
    \label{fig:mcs}
\end{figure}

Fig.~\ref{fig:mcs} summarizes the MCS sweep. MCS 0--3 form an
overhead-limited regime where \(T_p^\star\) remains at the 180 ms
upper bound, while MCS 4--9 enter an aging-limited regime where
\(T_p^\star\) rapidly decreases. Thus, MCS-aware adaptive sounding
improves throughput while preserving batteryless coexistence.

The two regimes are consistent with the analysis in Section~II. For
low MCS values, the Wi-Fi packet error probability remains low over
most of the 180 ms window, so the dominant penalty is the fixed
sounding overhead and longer cycles are preferred. For high MCS
values, the link operates closer to its SINR threshold; channel aging
and passive-overlay distortion jointly reduce the marginal value of
late packets, so frequent sounding becomes beneficial. The proposed
adaptive selection therefore avoids the inefficiency of a single
fixed sounding period across all MCS values.

\section{Conclusion}
\label{sec:conclusion}

In this paper, we investigated batteryless channel sounding for a
Wi-Fi 8-inspired DL MU-MIMO system. A packet-level cross-layer model
was developed to jointly capture periodic sounding overhead, CSI
aging, Wi-Fi packet reliability, and the additional throughput
provided by the Glaze-assisted passive overlay. Based on this model,
the post-sounding transmission interval \(T_{\mathrm p}\) was
optimized under different MCS and passive-overlay operating
conditions using a data-driven search method.The results show that the optimal sounding interval is determined by
the balance between sounding overhead and throughput degradation
caused by outdated CSI. At low MCS levels, the system is mainly
overhead-limited, and a longer \(T_{\mathrm p}\) is preferred. At
higher MCS levels, channel aging becomes dominant, resulting in a
shorter optimal sounding interval. The passive overlay can further
shift \(T_{\mathrm p}^{\star}\) by introducing both additional
batteryless throughput and Wi-Fi reliability loss. These findings
demonstrate that fixed-period sounding is insufficient for
passive-enabled MU-MIMO WLANs. Future work will consider joint
adaptation of \(T_{\mathrm p}\), attenuation depth, and passive
embedding rate, together with experimental validation on an SDR
testbed.

\bibliographystyle{IEEEtran}
\bibliography{ref}

@article{zhang2025crosslayer,
  author={Zhang, Lyutianyang and Cao, Liu and Wei, Dongyu and Chen, Mingzhe and Chen, Zhengchuan and Cui, Shuguang},
  title={Cross-Layer Channel Sounding Optimization Towards Next-Gen Wi-Fi: From Model Driven to Data Driven},
  journal={IEEE Transactions on Wireless Communications},
  year={2025},
  note={Early access},
  doi={10.1109/TWC.2025.3603182}
}

@article{galati2024wifi8,
  author={Galati Giordano, Lorenzo and Geraci, Giovanni and Carrascosa, Marc and Bellalta, Boris},
  title={What Will Wi-Fi 8 Be? A Primer on IEEE 802.11bn Ultra High Reliability},
  journal={IEEE Communications Magazine},
  volume={62},
  number={8},
  pages={126--132},
  year={2024},
  doi={10.1109/MCOM.001.2300728}
}

@standard{ieee80211ax2021,
  title={{IEEE} Standard for Information Technology--Telecommunications and Information Exchange Between Systems Local and Metropolitan Area Networks--Specific Requirements Part 11: Wireless LAN Medium Access Control and Physical Layer Specifications Amendment 1: Enhancements for High-Efficiency WLAN},
  organization={IEEE},
  number={IEEE Std 802.11ax-2021},
  year={2021},
  doi={10.1109/IEEESTD.2021.9442429}
}

@article{khorov2019tutorial,
  author={Khorov, Evgeny and Kiryanov, Anton and Lyakhov, Andrey and Bianchi, Giuseppe},
  title={A Tutorial on IEEE 802.11ax High Efficiency WLANs},
  journal={IEEE Communications Surveys \& Tutorials},
  volume={21},
  number={1},
  pages={197--216},
  year={2019},
  doi={10.1109/COMST.2018.2871099}
}

@inproceedings{zhang2023sounding,
  author={Zhang, Jingyuan and Avallone, Stefano and Blough, Douglas M.},
  title={Implementation and Evaluation of IEEE 802.11ax Channel Sounding Frame Exchange in ns-3},
  booktitle={Proceedings of the 2023 Workshop on ns-3},
  pages={10--18},
  year={2023},
  doi={10.1145/3592149.3592152}
}

@article{sangdeh2021deepmux,
  author={Sangdeh, Pedram Kheirkhah and Zeng, Huacheng},
  title={DeepMux: Deep-Learning-Based Channel Sounding and Resource Allocation for IEEE 802.11ax},
  journal={IEEE Journal on Selected Areas in Communications},
  volume={39},
  number={8},
  pages={2333--2346},
  year={2021},
  doi={10.1109/JSAC.2021.3087246}
}

@article{truong2013aging,
  author={Truong, Kien T. and Heath, Robert W.},
  title={Effects of Channel Aging in Massive MIMO Systems},
  journal={Journal of Communications and Networks},
  volume={15},
  number={4},
  pages={338--351},
  year={2013},
  doi={10.1109/JCN.2013.000065}
}

@article{kapetanovic2019glaze,
  author={Kapetanovic, Zerina and Saffari, Ali and Chandra, Ranveer and Smith, Joshua R.},
  title={Glaze: Overlaying Occupied Spectrum with Downlink IoT Transmissions},
  journal={Proceedings of the ACM on Interactive, Mobile, Wearable and Ubiquitous Technologies},
  volume={3},
  number={4},
  articleno={137},
  year={2019},
  doi={10.1145/3369825}
}

@inproceedings{liu2013ambient,
  author={Liu, Vincent and Parks, Aaron and Talla, Vamsi and Gollakota, Shyamnath and Wetherall, David and Smith, Joshua R.},
  title={Ambient Backscatter: Wireless Communication Out of Thin Air},
  booktitle={Proceedings of the ACM SIGCOMM 2013 Conference},
  pages={39--50},
  year={2013},
  doi={10.1145/2486001.2486015}
}

@inproceedings{kellogg2014wifibackscatter,
  author={Kellogg, Bryce and Parks, Aaron and Gollakota, Shyamnath and Smith, Joshua R. and Wetherall, David},
  title={Wi-Fi Backscatter: Internet Connectivity for RF-Powered Devices},
  booktitle={Proceedings of the ACM SIGCOMM 2014 Conference},
  pages={607--618},
  year={2014},
  doi={10.1145/2619239.2626319}
}

@misc{cao2025lightweightcoordinateconditioneddiffusionapproach,
      title={A Lightweight Coordinate-Conditioned Diffusion Approach for {6G C-V2X} Radio Environment Maps}, 
      author={Liu Cao and Zhaoyu Liu and Dongyu Wei and Yuan Yang and Yukun Pan and Lyutianyang Zhang},
      year={2025},
      eprint={2512.22535},
      archivePrefix={arXiv},
      primaryClass={cs.NI},
      url={https://arxiv.org/abs/2512.22535}, 
}

@article{gu2025matthew,
  title={The Matthew effect of AI programming assistants: A hidden bias in software evolution},
  author={Gu, Fei and Liang, Zi and Ma, Jiahao and Li, Hongzong},
  journal={arXiv preprint arXiv:2509.23261},
  year={2025}
}

@article{cao2025revisiting,
  title={Revisiting downlink mu-mimo in wi-fi 7/8: Insights from a case study},
  author={Cao, Liu and Zhang, Lyutianyang and Roy, Sumit},
  journal={IEEE Communications Standards Magazine},
  year={2025},
  publisher={IEEE}
}

\end{document}